\documentclass[preprintnumbers,amsmath,amssymbm,prd]{revtex4}
\usepackage{epsfig}
\usepackage{graphicx}
\usepackage{amssymb}

\begin{document}
\title{No-go theorem for spatially regular boson stars made of static nonminimally coupled massive scalar fields}
\author{Shahar Hod}
\affiliation{The Ruppin Academic Center, Emeq Hefer 40250, Israel}
\affiliation{ }
\affiliation{The Hadassah Institute, Jerusalem 91010, Israel}
\date{\today}

\begin{abstract}
\ \ \ We present a compact theorem which reveals the fact that
static spatially regular massive scalar fields with nonminimal
coupling to gravity cannot form spherically symmetric asymptotically
flat horizonless matter configurations. In particular, the no-go
theorem rules out the existence of boson stars made of static scalar
fields with generic values of the physical parameter $\xi$ which
quantifies the coupling between the spacetime curvature and the
massive bosonic fields.
\end{abstract}
\bigskip
\maketitle

\section{Introduction}

The non-linearly coupled Einstein-massive-scalar field equations
have a physically interesting and mathematically elegant structure
that has attracted the attention of researches during the last five
decades (see
\cite{Fra,Lie,Cha,Bek1,Teit,Heu,Bek2,BekMay,Bek20,Hodnmc,Hodrc,Herkr}
and references therein). In particular, this physical system is
known to possess asymptotically flat bound-state solutions in the
form of rotating hairy black holes that support spatially regular
{\it stationary} massive scalar field configurations
\cite{Hodrc,Herkr}.

Intriguingly, the phenomenologically rich Einstein-massive-scalar
system also admits horizonless bound-state solutions known as boson
stars. These physically interesting compact objects are made of
spatially regular {\it stationary} massive scalar fields. The
physical and mathematical properties of these self-gravitating
horizonless matter configurations have been extensively explored by
physicists and mathematicians (see \cite{Fra,Lie} and references
therein). In particular, the intriguing possibility has been
considered in the physics literature that boson stars may contribute
to the elusive dark matter distributions in galaxies \cite{Fra,Lie}.
In addition, compact bound-state scalar configurations (boson stars)
have been considered in the literature as physically exotic
horizonless mimickers of classical black-hole spacetimes
\cite{Fra,Lie}.

The main goal of the present paper is to reveal the intriguing fact
that, as opposed to {\it stationary} matter fields, {\it static}
self-gravitating massive scalar fields with non-minimal coupling to
gravity {\it cannot} form spatially regular horizonless bound-state
matter configurations (boson stars). In this context, it is
interesting to note that the asymptotic analysis of the non-linearly
coupled Einstein-scalar field equations presented in \cite{Hodrs1}
can be used to rule out the existence of static boson stars
(spatially regular matter configurations) made of nonminimally
coupled massive scalar fields in the dimensionless regime $\xi<0$ of
the nonminimal coupling parameter \cite{Notexi}.

In the present paper we shall explore the physical and mathematical
properties of the Einstein-scalar equations in the complementary
physical regime $\xi\geq0$ of the dimensionless nonminimal coupling
parameter. In particular, below we shall present a remarkably
compact no-go theorem for static boson stars which reveals the fact
that the non-linearly coupled Einstein-massive-scalar field
equations do not admit static spatially regular bound-state
solutions made of self-gravitating scalar fields with nonminimal
coupling to gravity.

\section{Description of the system}

We consider a spatially regular asymptotically flat spacetime whose
curvature is sourced by a static self-gravitating massive scalar
field $\psi$ with nonminimal coupling to gravity. The curved
spherically symmetric spacetime is described by the line element
\cite{BekMay,Notecor,Noteunit}
\begin{equation}\label{Eq1}
ds^2=-e^{\nu}dt^2+e^{\lambda}dr^2+r^2(d\theta^2+\sin^2\theta
d\phi^2)\ ,
\end{equation}
where $\nu=\nu(r)$ and $\lambda=\lambda(r)$.

The metric functions describing the spatially regular matter
configurations are characterized by the near-origin functional
behaviors \cite{Hodt1}
\begin{equation}\label{Eq2}
e^{\lambda}(r\to 0)=1+O(r^2)\ \ \ \ ; \ \ \ \ 0<e^{\nu}<\infty\
\end{equation}
with \cite{Notetag}
\begin{equation}\label{Eq3}
\lambda'(r\to0)\to0\ \ \ \ ; \ \ \ \ \nu'(r\to0)\to0\  .
\end{equation}
In addition, the metric functions of the asymptotically flat
spacetime are characterized by the far-region ($r\to\infty$) functional relations
\cite{BekMay}
\begin{equation}\label{Eq4}
\nu(M/r\to0)\sim M/r\ \ \ \ ; \ \ \ \ \lambda(M/r\to0)\sim
M/r\ ,
\end{equation}
where $M$ is the asymptotically measured finite ADM mass of the
spatially regular static matter configuration.

The self-gravitating scalar field is characterized by the action
\cite{BekMay,NoteSEH}
\begin{equation}\label{Eq5}
S=S_{EH}-{1\over2}\int\big(\partial_{\alpha}\psi\partial^{\alpha}\psi+\mu^2\psi^2+\xi
R\psi^2\big)\sqrt{-g}d^4x\ ,
\end{equation}
where $\mu$ \cite{Notemu} is the mass of the matter field and the
dimensionless nonminimal coupling parameter $\xi$ quantifies the
coupling between the scalar curvature $R(r)$ [see Eq. (\ref{Eq22})
below] of the spatially regular spacetime and the massive scalar
field $\psi$. For later purposes we note that asymptotically flat
spacetimes are characterized by the far-region functional behavior
\cite{BekMay}
\begin{equation}\label{Eq6}
R(M/r\to0)\to 0\
\end{equation}
of the scalar curvature. In addition, spatially regular nonminimally
coupled scalar matter configurations are characterized by
asymptotically finite radial eigenfunctions. In particular, as
discussed in \cite{BekMay}, the bounded functional relation
\begin{equation}\label{Eq7}
-\infty<8\pi\xi\psi^2<1\ \ \ \ \text{for}\ \ \ \ M/r\ll1\  .
\end{equation}
for the scalar eigenfunction guarantees that the corresponding
physically acceptable curved spacetimes are characterized by finite
and positive values of the asymptotic effective gravitational
constant $G_{\text{eff}}=G(1-8\pi G\xi\psi^2)$ \cite{BekMay}.

The spatial and temporal functional behaviors of the nonminimally
coupled massive scalar fields with the action (\ref{Eq5}) are
governed by the compact partial differential equation \cite{BekMay}
\begin{equation}\label{Eq8}
\partial_{\alpha}\partial^{\alpha}\psi-(\mu^2+\xi R)\psi=0\  .
\end{equation}
In particular, for spherically symmetric static matter
configurations, one obtains from (\ref{Eq1}) and (\ref{Eq8}) the
characteristic ordinary differential equation
\begin{equation}\label{Eq9}
\psi{''}+{1\over2}\big({{4}\over{r}}+\nu{'}-\lambda{'}\big)\psi{'}-(\mu^2+\xi
R) e^{\lambda}\psi=0\  ,
\end{equation}
which determines the spatial behavior of the scalar eigenfunction in
the curved spacetime (\ref{Eq1}).

The components of the energy-momentum tensor $T^{\alpha}_{\beta}$,
which characterizes the nonminimally coupled static scalar fields
with the action (\ref{Eq5}), are given by the functional expressions
\cite{BekMay}
\begin{equation}\label{Eq10}
T^{t}_{t}=e^{-\lambda}{{\xi(4/r-\lambda{'})\psi\psi{'}+(2\xi-1/2)(\psi{'})^2+2\xi\psi\psi{''}}
\over{1-8\pi\xi\psi^2}}-{{\mu^2\psi^2}\over{2(1-8\pi\xi\psi^2)}}\ ,
\end{equation}
\begin{equation}\label{Eq11}
T^{r}_{r}=e^{-\lambda}{{\xi(4/r+\nu{'})\psi\psi{'}+(\psi{'})^2/2}
\over{1-8\pi\xi\psi^2}}-{{\mu^2\psi^2}\over{2(1-8\pi\xi\psi^2)}}\ ,
\end{equation}
and
\begin{equation}\label{Eq12}
T^{\theta}_{\theta}=T^{\phi}_{\phi}=e^{-\lambda}{{\xi(2/r+\nu{'}-\lambda{'})\psi\psi{'}+(2\xi-1/2)(\psi{'})^2+2\xi\psi\psi{''}}
\over{1-8\pi\xi\psi^2}}-{{\mu^2\psi^2}\over{2(1-8\pi\xi\psi^2)}}\  .
\end{equation}
As explicitly proved in \cite{BekMay}, physically acceptable spacetimes are
characterized by finite components of the energy-momentum tensor:
\begin{equation}\label{Eq13}
\{|T^{t}_{t}|,|T^{r}_{r}|,|T^{\theta}_{\theta}|,|T^{\phi}_{\phi}|\}<\infty\  .
\end{equation}

\section{The no-go theorem for the static nonminimally coupled
massive scalar field configurations (boson stars) in the regime
$\xi\geq0$}

As emphasized above, the asymptotic ($r\gg M$) analysis of the
Einstein-scalar field equations presented in \cite{Hodrs1,Notegib}
rules out the existence of static boson stars made of
self-gravitating massive scalar fields in the dimensionless regime
$\xi<0$ of the nonminimal coupling parameter. In the present section
we shall complete the no-go theorem for the static sector of
horizonless boson stars. In particular, we shall explicitly prove
that nonminimally coupled massive scalar fields with a dimensionless
field-curvature coupling parameter in the physical regime $\xi\geq0$
cannot form spatially regular asymptotically flat static matter
configurations.

To this end, we first note that the radial scalar equation
(\ref{Eq9}) takes the simple asymptotic form [see Eqs. (\ref{Eq4}) and
(\ref{Eq6})] \cite{Hodrs1}
\begin{equation}\label{Eq14}
\psi{''}+{{2}\over{r}}\psi{'}-\mu^2\psi=0\
\end{equation}
in the $M/r\ll1$ region. The physically acceptable (normalizable)
radial solution of (\ref{Eq14}) which respects the bounded asymptotic
behavior (\ref{Eq7}) of physically acceptable spacetimes \cite{BekMay} is given by
\begin{equation}\label{Eq15}
\psi(r)=A\cdot{{e^{-\mu r}}\over{\mu r}}\ \ \ \ \text{for}\ \ \ \
M/r\ll1\ ,
\end{equation}
where $A$ is a normalization constant.

We shall now prove that the eigenfunction $\psi(r)$, which
characterizes the spatial behavior of the spherically symmetric
static scalar field configurations, has a non-monotonic dependence on the areal
coordinate $r$. In particular, if the radial scalar eigenfunction is
characterized by the near-origin relation $\psi(r\to0)\to0$, then
its asymptotic behavior [see Eq. (\ref{Eq15})]
\begin{equation}\label{Eq16}
\psi(M/r\to0)\to0\
\end{equation}
would immediately imply that it must possess (at least) one extremum
point $r_{\text{peak}}\in(0,\infty)$ in the curved spacetime (\ref{Eq1}). It therefore remains to be
proved that nonminimally coupled massive scalar fields which are
characterized by the near-origin behavior
\begin{equation}\label{Eq17}
\psi(r\to0)\neq0\
\end{equation}
must also posses an extremum point.

For later purposes, we note that one finds from the energy-momentum
components (\ref{Eq10}) and (\ref{Eq12}) the simple functional relation
\begin{equation}\label{Eq18}
T^{t}_{t}-T^{\phi}_{\phi}=e^{-\lambda}{{\xi(2/r-\nu{'})\psi\psi{'}}
\over{1-8\pi\xi\psi^2}}\  .
\end{equation}
In particular, taking cognizance of Eqs. (\ref{Eq2}), (\ref{Eq3}), (\ref{Eq13}), and (\ref{Eq18}),
one deduces the characteristic near-origin functional behavior
\begin{equation}\label{Eq19}
|{{1}\over{r}}\psi\psi{'}|<\infty\ \ \ \ \text{for}\ \ \ \ r\to0
\end{equation}
for spatially regular scalar matter configurations.
Likewise, taking cognizance of Eqs. (\ref{Eq2}), (\ref{Eq3}), (\ref{Eq13}), (\ref{Eq17}), (\ref{Eq19}),
and the relation [see Eqs. (\ref{Eq10}) and (\ref{Eq11})]
\begin{equation}\label{Eq20}
T^{t}_{t}-T^{r}_{r}=e^{-\lambda}{{(2\xi-1)(\psi{'})^2+2\xi\psi\psi{''}-\xi(\nu+\lambda){'}\psi\psi{'}}
\over{1-8\pi\xi\psi^2}}\  ,
\end{equation}
one finds the bounded near-origin functional behavior
\begin{equation}\label{Eq21}
|\psi{''}|<\infty\ \ \ \ \text{for}\ \ \ \ r\to0\
\end{equation}
for spatially regular static field configurations.

Substituting the Ricci scalar curvature \cite{Noteric}
\begin{equation}\label{Eq22}
R=-{{8\pi}\over{1-8\pi\xi\psi^2}}\Big\{e^{-\lambda}\Big[\xi\big({{12}\over{r}}+3\nu{'}-3\lambda{'}\big)\psi\psi{'}+
6\xi\psi\psi{''}+(6\xi-1)(\psi{'})^2\Big]-2\mu^2\psi^2\Big\}\
\end{equation}
which characterizes the spherically symmetric self-gravitating massive scalar fields
into Eq. (\ref{Eq9}), one obtains the ordinary differential equation
\begin{equation}\label{Eq23}
{\cal F}\cdot\psi{''}
+\psi{'}\cdot\Big[{1\over2}\big({{4}\over{r}}+\nu{'}-\lambda{'}\big){\cal F}
+8\pi\xi(6\xi-1)\psi\psi{'}\Big]
-\mu^2e^{\lambda}\big(1+8\pi\xi\psi^2\big)\psi=0\  ,
\end{equation}
which determines the radial behavior of the nonminimally coupled scalar matter configurations, where
\begin{equation}\label{Eq24}
{\cal F}(r;\xi)\equiv 1+8\pi\xi(6\xi-1)\psi^2\  .
\end{equation}
For later purposes we note that, taking cognizance of Eq. (\ref{Eq16}),
one finds that ${\cal F}(r;\xi)$ is characterized by the simple
asymptotic behavior
\begin{equation}\label{Eq25}
{\cal F}(M/r\to0)\to1\  .
\end{equation}

The radially-dependent function ${\cal F}(r;\xi)$ is obviously
positive definite for nonminimally coupled massive scalar fields
in the dimensionless regimes $\xi\geq 1/6$ and $\xi\leq0$ [see Eq. (\ref{Eq24})].
We shall now prove that this property also characterizes
the function ${\cal F}(r)$ in the complementary dimensionless regime $0<\xi<1/6$ of the
nonminimal coupling parameter.
Let us assume that the function ${\cal F}(r)$ has a root at some
point $r_0\in(0,\infty)$, in which case one finds from (\ref{Eq23}) the functional relation
\cite{Notepno}
\begin{equation}\label{Eq26}
8\pi\xi(6\xi-1)(\psi{'})^2=\mu^2e^{\lambda}(1+8\pi\xi\psi^2)\ \ \ \
\text{at}\ \ \ \ r=r_0\  .
\end{equation}
Taking cognizance of the fact that, in the dimensionless
regime $0<\xi<1/6$, the l.h.s of (\ref{Eq26}) is non-positive
whereas the r.h.s of (\ref{Eq26}) is positive definite, one deduces that
the equality sign in (\ref{Eq23}) cannot be respected at the assumed
root $r=r_0$ of ${\cal F}(r)$.
We therefore learn that the radial function ${\cal F}(r)$ has no roots.
In particular, using the characteristic asymptotic relation (\ref{Eq25}),
one finds that the radially-dependent function ${\cal F}(r;\xi)$ is
characterized by the simple inequality
\begin{equation}\label{Eq27}
{\cal F}(r)>0\ \ \ \ \text{for}\ \ \ \ r\in(0,\infty)\
\end{equation}
in the entire curved spacetime.

Taking cognizance of Eqs. (\ref{Eq2}), (\ref{Eq3}), (\ref{Eq19}), (\ref{Eq21}),
and (\ref{Eq23}), one obtains the near-origin ($r\to0$) radial equation
\begin{equation}\label{Eq28}
\psi{''}+{{2}\over{r}}\psi{'}
-\mu^2{\cal F}^{-1}\big(1+8\pi\xi\psi^2\big)\psi=0\  ,
\end{equation}
whose physically acceptable solution \cite{Notepac} is characterized
by the small-$r$ spatial behavior \cite{Notefoh,Noteaq0}
\begin{equation}\label{Eq29}
\psi(r\to0)=B\Big[1+{1\over6}\mu^2{{1+8\pi\xi
B^2}\over{1+8\pi\xi(6\xi-1)B^2}}\cdot r^2\Big]+O(r^3)\  ,
\end{equation}
where $B$ is a constant. Interestingly, and most importantly for our
analysis, one learns from Eqs. (\ref{Eq24}), (\ref{Eq27}), and
(\ref{Eq29}) that spatially regular nonminimally coupled massive
scalar fields with $\xi\geq0$ are characterized by the near-origin
functional relations
\begin{equation}\label{Eq30}
\psi\psi'(r\to0)=0\ \ \ \ ; \ \ \ \ \psi\psi''(r\to0)>0\ .
\end{equation}

The asymptotic large-$r$ and small-$r$ functional behaviors
(\ref{Eq16}) and (\ref{Eq30}), when combined together, reveal the important fact that the
characteristic radial eigenfunction $\psi(r)$ of the spatially
regular nonminimally coupled massive scalar fields must possess, in
accord with our previous assertion, (at least) one extremum point
$r_{\text{peak}}\in(0,\infty)$ in the curved spacetime (\ref{Eq1}). In particular, at this extremum
point the scalar eigenfunction is characterized by the functional
relations
\begin{equation}\label{Eq31}
\{\psi\neq0\ \ \ ; \ \ \ \psi{'}=0\ \ \ ; \ \ \
\psi\cdot\psi{''}<0\}\ \ \ \ \text{for}\ \ \ \ r=r_{\text{peak}}\  .
\end{equation}

Substituting (\ref{Eq31}) into the radial differential equation
(\ref{Eq23}), one finds
\begin{equation}\label{Eq32}
{\cal F}\cdot\psi\psi{''}=\mu^2e^{\lambda}(1+8\pi\xi\psi^2)\psi^2\ \
\ \ \text{at}\ \ \ \ r=r_{\text{peak}}\  .
\end{equation}
Interestingly, taking cognizance of the analytically derived
relations (\ref{Eq27}) and (\ref{Eq31}) one learns that, in the
dimensionless physical regime $\xi\geq0$, the l.h.s of (\ref{Eq32})
is negative definite whereas the r.h.s of (\ref{Eq32}) is positive
definite. Thus, for spatially regular static scalar configurations,
the radial differential equation (\ref{Eq23}) cannot be satisfied at
the characteristic extremum point (\ref{Eq31}).

We therefore conclude that static spatially regular massive scalar
fields whose nonminimal coupling parameter lies in the
dimensionless physical regime $\xi\geq0$ cannot form spherically symmetric asymptotically
flat horizonless matter configurations (boson stars).

\section{Summary}

Boson stars represent self-gravitating horizonless compact
objects which are made of massive scalar fields. The
physical and mathematical properties of these spatially
regular bound-state matter configurations have attracted
the attention of physicists and mathematicians during the last three decades
(see \cite{Fra,Lie} and references therein).

In the present paper we have presented a novel non-existence theorem for static
boson stars. In particular, the static sector of the non-linearly coupled
Einstein-matter field equations has been studied analytically for spatially
regular self-gravitating massive scalar fields with nonminimal coupling to
gravity. Interestingly, our compact no-go theorem explicitly rules out the
existence of asymptotically flat horizonless spacetimes that contain
spatially regular boson stars which are made of nonminimally coupled
\cite{Notexip} massive scalar fields.

\bigskip
\noindent
{\bf ACKNOWLEDGMENTS}
\bigskip

This research is supported by the Carmel Science Foundation. I would
like to thank Yael Oren, Arbel M. Ongo, Ayelet B. Lata, and Alona B.
Tea for helpful discussions.


\begin{thebibliography}{99}
\bibitem{Notecor} We use the familiar Schwarzschild spacetime coordinates $(t,r,\theta,\phi)$.
\bibitem{Noteunit} Natural units in which $G=c=\hbar=1$ are used throughout the paper.
\bibitem{Notetag} We use the prime symbol ${'}$ to denote a radial derivative with respect to
the areal coordinate $r$.
\bibitem{NoteSEH} Here $S_{\text{EH}}$ is the Einstein-Hilbert action of general relativity.
\bibitem{Notemu} Note that the physical parameter $\mu$ in the action (\ref{Eq5})
stands for $\mu/\hbar$. This mass parameter of the nonminimally
coupled scalar field therefore has the dimensions of
(length)$^{-1}$.
\bibitem{Noteric} Here we have used the expressions (\ref{Eq10}), (\ref{Eq11}), and
(\ref{Eq12}) for the components of the energy-momentum tensor
together with the Einstein relation $R=-8\pi T$ between the Ricci scalar curvature and the
trace of the corresponding energy-momentum tensor.
\bibitem{Notepno} Note that one finds from Eq. (\ref{Eq24})
the trivial relation $\psi(r=r_0)\neq0$ at the assumed root of ${\cal F}$.
\bibitem{Notepac} That is, the mathematical solution of the radial differential
equation (\ref{Eq28}) which respects the characteristic near-origin functional
relations (\ref{Eq19}) and (\ref{Eq21}) for spatially regular scalar eigenfunctions.
\bibitem{Notefoh} It is worth emphasizing that one
learns from Eqs. (\ref{Eq17}), (\ref{Eq19}), (\ref{Eq21}), and (\ref{Eq29})
that ${\cal F}(r)$ cannot vanish in the near-origin $r\to0$ limit.
\bibitem{Noteaq0} Note that $B\equiv\psi(r\to0)\neq0$ [see Eqs. (\ref{Eq17}) and (\ref{Eq29})].
As discussed above, if the scalar eigenfunction
is characterized by the near-origin relation $\psi(r\to0)=0$,
then the asymptotic radial behavior (\ref{Eq16}) implies,
in accord with the above stated assertion, that $\psi(r)$
is a non-monotonic function.
\bibitem{Notexip} It is worth noting again that our compact no-go
theorem is valid for non-negative $\xi\geq0$ values of the dimensionless
physical parameter $\xi$ which quantifies the coupling between the massive
scalar field and the characteristic curvature of the spherically
symmetric static spacetime [see Eq. (\ref{Eq5})]. As emphasized above,
the non-existence of spatially regular static boson stars made of
nonminimally coupled massive scalar fields in the complementary
regime $\xi<0$ of the dimensionless nonminimal coupling parameter $\xi$
can be deduced from the large-$r$ asymptotic analysis presented in \cite{Hodrs1}.
\end{thebibliography}
\end{document}